# Curing and post-curing luminescence in an epoxy resin

O. Gallot-lavallée [1]*, G. Teyssedre [1], C. Laurent [1], S. Robiani [2] and S. Rowe [2]

[1] Université Paul Sabatier, Laboratoire de Génie Electrique de Toulouse,

118, route de Narbonne, Toulouse, 31062, France

[2] Schneider Electric, Direction des Recherches Matériaux,

Rue Henri Tarze, Grenoble, 38050, France

* E-mail : gallot-lavallee@lget.ups-tlse.fr

**Abstract**

A spontaneous luminescence is reported when epoxy resin samples are heated in air. This phenomenon is very sensitive to the nature of the atmosphere. The same treatment in nitrogen leads to an extinction of the luminescence. The emission process is restored when samples are kept for a sufficient time in air. In order to better understand this phenomenon, we have investigated the luminescence of the elementary constituents of the epoxy (resin and hardener) when heated in air and nitrogen, as well as during resin curing in the same atmospheres. It appears that the emission process is linked with the presence of oxygen. Although the kinetics of the luminescence can differ depending on the nature of the sample (cured resin, resin during curing, liquid components), the emission spectra are the same during resin curing and upon heating of the cured resin and hardener. The emission spectrum of the base resin is different. It is concluded that the light results from a chemiluminescence process during oxidation.

## 1 INTRODUCTION

Epoxy resins are widely used as electrical insulation in electrical engineering applications such as transformers, bus bars, electrical machines and so on [1]. However, and in spite of the fact that it is in no way a new material, very little is known about electrical ageing. Published reports have focussed on the effect of partial discharges and electrical treeing [2, 3] but there is comparatively scarce data on the possible influence of internal space charges generated by the thermo-electric stress itself. We have undertaken such a study by means of space charge, conduction and electroluminescence measurements with the objective of correlating space charge accumulation and electroluminescence emission considered as an ageing indicator [4]. A step along this approach was to characterize the photo-luminescence properties of the epoxy in order to provide a background for the discussion of the nature of the excited states involved when applying a voltage to the resin [5]. Indeed, as usual when dealing with luminescence of polymeric materials, the interpretation of emission spectra in terms of molecular species is by no way a straightforward task. This is particularly true when considering epoxies since luminescence of the cured resin may arise from the complex and often only partially documented formulation of the main compounds, hardener and base resin, as well as from molecular structures formed during the curing itself. It is therefore advisable to have as much information as possible on the photo-physical and photo-chemical behaviour of the resin for further understanding of its potential evolution under an electric field.

During this investigation, we realized that our epoxy discs emitted a spontaneous light when heated. A similar effect has been reported very recently by Suzuki et al. [6]. They attributed this light emission upon heating of cured samples to chemiluminescence (CL), suggesting that the light emission proceeds by a two step process. The first step is the formation of an unspecified substance causing CL and the second step is its reaction producing luminescence emission. The first step is predominant at room temperature, so that the CL-causing substance accumulates when a specimen is left in ambient atmosphere. By heating the specimen, CL is observed. In an inert atmosphere, the total amount of CL is in a linear relation with the square root of time during which the specimen is left in an ambient atmosphere pointing towards a diffusion-controlled generation of the CL-causing substance. During heating in an inert atmosphere the luminescence is exhausted because the CL-causing substance cannot be regenerated. The authors proposed that oxygen could be the initiator for the formation of the CL-causing substance.

The aim of this study is to discuss this phenomenon by bringing additional information. Measurements performed in air and nitrogen on cured epoxy samples support the observation by Suzuki et al. Furthermore, the same experiment is performed on the epoxy components (resin and hardener). The spectral analysis of the luminescence during curing of the resin is also reported, as well as the emission spectra during heating of cured epoxy, hardener and base resin.

## 2 EXPERIMENTAL

### 2.1 Materials

The investigated resin was produced by mixing an equal weight of hardener HY227 and resin CY225, hereafter referred to as base resin, and a small quantity of catalyser DY062 (Huntsman's references). The known chemical properties of the components are given in Table 1, where DGEBA stands for Diglycidyl ether of bisphenol A and MTHPA means Methyltetrahydrophtalic anhydride. An epoxy resin with a relatively low glass transition temperature (Tg) of 65°C is obtained, due to the presence of a plasticizer in the hardener. This plasticizer builds longer molecule between crosslinking.

A first kind of experiment was performed on cured epoxy resin. Hardener, base resin and catalyser were introduced in a mould and mixed for 15 min at 60 °C under primary vacuum in order to outgas products. Then, curing was achieved at 100 °C for 16 h providing disks of 500 µm (±10µm) in thickness and 110 mm in diameter.

A second kind of experiment was performed on the separate components (base resin, hardener) in their viscous state, or during curing of a base resin-hardener mixture (no added catalyser). A quartz crucible was used for this purpose.

### 2.2 Experimental set-up

The home-made chamber for luminescence experiments is schematically described in Figure 1. Samples are disposed on a reservoir allowing temperature control from 100 K to 400 K. A cooled photomultiplier (PM) working in photon counting mode is used for the luminescence measurement. An optical lens focuses the light onto the photocathode of the PM. Its spectral response is wide and almost flat from 200 nm to 800 nm allowing detection in the whole visible range. The luminescence spectra were realized through the use of a Charged Coupled Device (CCD) camera (detection range from 190 nm to 820 nm) coupled to a monochromator (resolution of 3 nm). The light from the sample is focussed onto the entrance slit of the monochromator. The spectra reported in this paper have not been corrected by the wavelength dependence of the response of the camera and monochromator. These corrections have however a minor impact on the spectral shape when working between 300 nm to 700 nm which is the case here.

The temperature was increased linearly at a rate of 5°C/min up to the target level. Thermal regulation is effective through a feed back by a sensor mounted on the sample holder. In these conditions, reasonable temperature control and homogeneity within the sample is obtained, even so there may be some temperature gradient in the system. Experiments were carried out either in air or nitrogen. In the latter case, the chamber is evacuated for 12 hours under a secondary vacuum ($10^{-6}$ mb) prior to introducing an atmospheric pressure of nitrogen. Specific experimental conditions are specified when necessary.

## 3 RESULTS

### 3.1 Luminescence upon heating a cured sample

Figure 2 shows the spontaneous luminescence (SL) measured on a cured sample under a nitrogen atmosphere by rising the temperature from 25 to 72°C and keeping it constant. Light is maximum when the temperature reaches 72°C. It decreases exponentially for about 30 min. After this, more gradual decrease is observed until it disappears completely after about four hours.

A subsequent series of experiments were performed according to the following scheme:

1/ repeating the same experiment as above, at 72°C in nitrogen, and waiting for the emitted light to reach the photomultiplier noise level;

2/ cooling down the sample to room temperature, waiting for 48h, and repeating the experiment in the same conditions as in 1/, the initial nitrogen atmosphere being unchanged;

3/ cooling down again the sample to room temperature, exposing it to air for 48 h, and re-heating the sample to 72 °C, in air.

Results obtained for cycles 1/ and 3/ are shown in Figure 3. No light was detected in step 2/ as seen in Figure 4. The light intensity in cycles 1/ and 3/ was larger compared to the results of Figure 3 because the analyzed sample area was larger. Clearly, the nature of the atmosphere has a strong effect on the light being detected even if the dynamic in both experiments seems to be similar in short term. It has to be emphasized that the peak of SL is not related to an overshoot in the sample temperature since heating is effective through the circulation of a liquid into the reservoir. In step 1/, i.e. in nitrogen, SL is continuously decreasing (except for the time interval where the temperature is increased). In step 3/ a quasi steady-state level of light is measured for a time > 1000 s. The air seems to provide the conditions for a steady state emission of the epoxy resin. The first hypothesis that comes to mind is a possible oxidation reaction of the resin; the SL in nitrogen might decrease due to consumption of oxygen by the reaction itself.

This behaviour is consistent with the report by Suzuki et al [6]. One new issue to help our understanding of this phenomenon was to record the SL spectrum during heating. It is shown in Figure 5. The spectrum, which is broad and structure-less, peaks at 562 nm. This is the domain of the phosphorescence emission of the cured resin as seen in photoluminescence experiments [5].

In order to better understand the origin of the luminescence observed during the heating up of the cured resin, we recorded the SL of the resin components in the liquid phase upon thermo-stimulation, and during resin curing.

### 3.2 Luminescence upon heating resin components and during resin curing

The liquid components (base resin and hardener) were degassed for 15 min at a pressure of 4 mb. For curing, the base resin-hardener mixture was stirred for 15 min at the same reduced pressure. Then, the liquid phases were transferred to the chamber under a nitrogen atmosphere before being heated up to 100 °C. No light was detected in these experimental conditions. Then, nitrogen was evacuated from the chamber and air was injected under atmospheric pressure. As soon as air entered the chamber, light emission increased quickly and stabilized afterwards. It appears obvious that the presence of oxygen is needed to excite SL, both in separate resin components and during resin curing. The same behaviour was observed when oxygen was introduced instead of air (which contains impurities and water), demonstrating that oxygen alone, combined with temperature, can monitor spontaneous light emission.

Table II gives a quantification of the emission processes relevant to both the liquid components and the resin during curing, when heated in air. It can be seen that the level of light being measured is more important in the hardener or hardener-base resin combination than in the base resin. In the course of these experiments, the emission was strong enough so as to allow spectral analysis to be carried out. Emission spectra recorded after 30min of heating are shown in Figure 6-a to c. For the hardener and the resin under curing, spectra are similar both in intensity and wavelength position. The spectrum is also qualitatively the same as the one obtained on a previously cured resin, Figure 5. As regards the spectrum of the base resin, it is blue shifted to 415nm, thereby showing that different species are involved in the emission. The spectrum is of lower intensity, consistently with data provided by PM counting, Table II. It is therefore an expected result that light emission from the hardener-base resin mixture be dominated by emission from the hardener.

## 4 DISCUSSION

An expected mechanism for light emission during resin curing is chemiluminescence resulting from direct radiative relaxation of excited states of species formed during curing, or energy transfer from excited states formed during curing to other species that decay radiatively afterwards. As no light was detected when resin components were degassed and put in an inert atmosphere prior to heating, the involvement of such mechanisms in the results presented here can be discarded. In addition, the intensity of light being measured with hardener alone and hardener mixed with base resin are of the same order, demonstrating that curing itself does not provide a significant contribution to light emission.

Since oxygen appears as the necessary and sufficient condition, combined with heating, to observe SL from all the materials being considered here, namely base resin, hardener, resin in the course of curing and previously cured resin, it can be concluded that what is observed is most likely chemiluminescence due to oxidation reactions. The spectral analysis tells us that the same species are responsible for the light emission during curing and upon heating of the hardener and the cured resin, but not the base resin. This is not surprising taking into account the chemical nature of the constituents. The hardener is incorporated into the cured resin and can therefore be responsible for the SL observed in the cured resin and during curing. These observations appear to be in line with reports by Le Huy & al. [7] indicating that in the temperature range 140-200°C, the ageing of anhydride cured epoxies is largely dominated by oxidation processes involving various reaction pathways of the anhydride groups. A distinct process of oxidation is likely to take place in the base resin leading to a peculiar emission spectrum in relation with the chemical formulation of the resin.

Turning back to spontaneous luminescence of previously cured resins, a post-curing effect can be reasonably discarded. Although the corresponding emission spectrum has the same shape as the one during curing of the mixture in air. In all the above SL observation, the presence of oxygen is necessary. This would seem to point towards an oxygen mechanism. Outgassing solid samples at room temperature appears not sufficient for removing oxygen dissolved in it so that transient luminescence is observed when heating such samples in nitrogen. Luminescence vanishes when dissolved oxygen has reacted or has left the sample under the effect of temperature. The level of spontaneous light as measured by Suzuki & al. [6] as a function of storage time at room temperature appears diffusion controlled (as is thermal oxidation [8]), just because oxygen uptake is diffusion controlled. However, on the basis of the present results, a simpler mechanism than that proposed in [6] can be envisaged, without recourse to the formation of unspecified CL causing substance. What happens at room temperature would be just diffusion-controlled oxygen uptake in cured or liquid samples. This can be easily removed when stirring liquid substances under vacuum, but much less efficiently in solid samples. Upon heating, oxygen would react with anhydride groups of the hardener, giving rise to SL, of transient shape in inert atmosphere because the available oxygen quantity is only that which is dissolved in the material and of steady state shape in air because oxygen is provided continuously by the environment. According to the studies lead by Le Huy & al. [7, 8], this oxidation was attested by the build-up of IR bands typically representative of oxygenated structures. It proceeds with weight loss, a decrease of the flexure and tensile strength, and a change of various physical properties including dielectric strength or Tg. As for the majority of industrial thermosets, post-crosslinking can predominate in the early period of exposure. This oxidation favours anhydrides monomer reformation, presumably from monoacid dangling groups.

## 5 CONCLUSIONS

Summarizing the results reported in this work:

-All the spontaneous luminescence processes observed during thermo-stimulation, being from cured resin, hardener, base resin and during resin curing are linked with the presence of oxygen,

-The SL can be regenerated by sample exposure to air at room temperature,

-The hardener seems to be responsible for SL in the cured resin as well as in liquid samples containing it.

It can therefore be concluded that chemiluminescence due to oxidation is responsible for all the observations. Thermal oxidation is clearly a possibility to consider when evaluating ageing of electrical devices containing such epoxy resin as insulation, even at moderate working temperature.

## Acknowledgments

Thanks are due to Francois Trichon, Yvan Sremin and Giovani Barbera; they facilitated the study of this epoxy resin. Thanks also to Ronan Healy.

*PS: Style realised by J Appl Polym Sci style from End Note*

## CAPTIONS

**Table 1**: Known chemical formulation of components of the resin. 10% of the hardener contents are unknown.

**Table 2**: Quantification of the luminescence emitted during heating the resin components and during resin curing, at 100 °C. Cps stands for photomultiplier counts per second.

**Figure 1**: Schematic of the experimental set-up for light detection

**Figure 2**: Luminescence kinetics measured on a cured sample under a nitrogen atmosphere (temperature rise of 5°C/min up to 72 °C). Cps stands for photomultiplier counts per second.

**Figure 3**: Comparison between the luminescence kinetics measured on a cured sample under both nitrogen and air atmospheres (temperature rise of 5°C/min up to 72 °C)

**Figure 4**: Effect of temperature cycling on the luminescence emitted by a cured resin in nitrogen (rate of temperature change is 5°C/min).

**Figure 5**: Luminescence spectrum of cured resin at 100°C

**Figure 6**: Luminescence spectra after 30 min at 100°C °C: (a)-resin-hardener mixture during curing, (b)-hardener, (c)-base resin.

## TABLES

| *Compound* | *Formulation* |
|---|---|
| *Base resin*<br>*(CY225)*<br>*100wt. parts* | 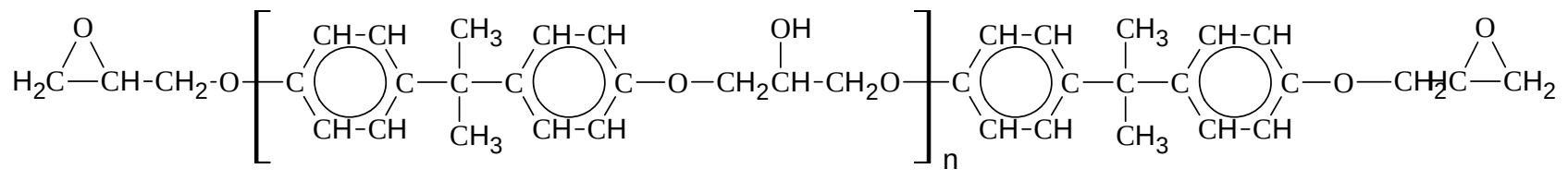<br>DGEBA n=1: 75wt.% + DGEBA n=2: 15wt.%<br>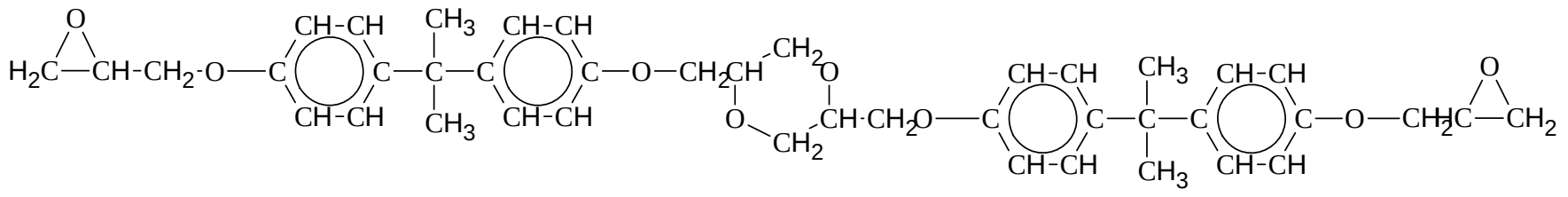<br>Homopolymer from DGEBA n=0: 9wt.% |
| *Hardener*<br>*(HY227)*<br>*100wt. parts* | 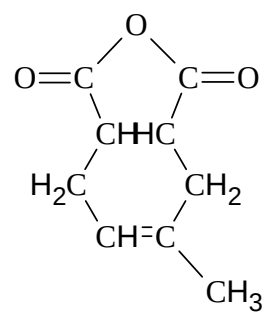<br>MTHPA: 50wt.%<br>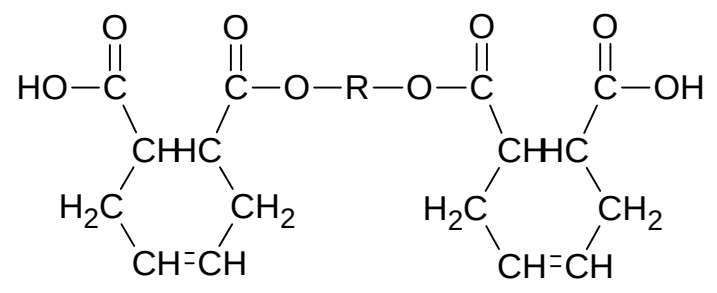<br>Plasticizer: di-ester di-carboxylic acid: 40wt.%<br>R is an aliphatic structure |
| *Catalyser*<br>*(DY062)*<br>*0.6wt. parts* | 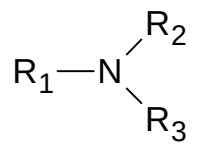<br>tertiary amine: 100% |

Table 1

*Known chemical formulation of components of the resin. 10% of the hardener contents are unknown.*

| Time (s) | 30 min | 50 min |
|---|---|---|
| Base resin | 150 cps | 300 cps |
| Hardener | 800 cps | 450 cps |
| Base resin-hardener mixture | 820 cps | 650 cps |

Table 2

*Quantification of the luminescence emitted during heating the resin components and during resin curing, at 100 °C. Cps stands for photomultiplier counts per second.*

**FIGURES**

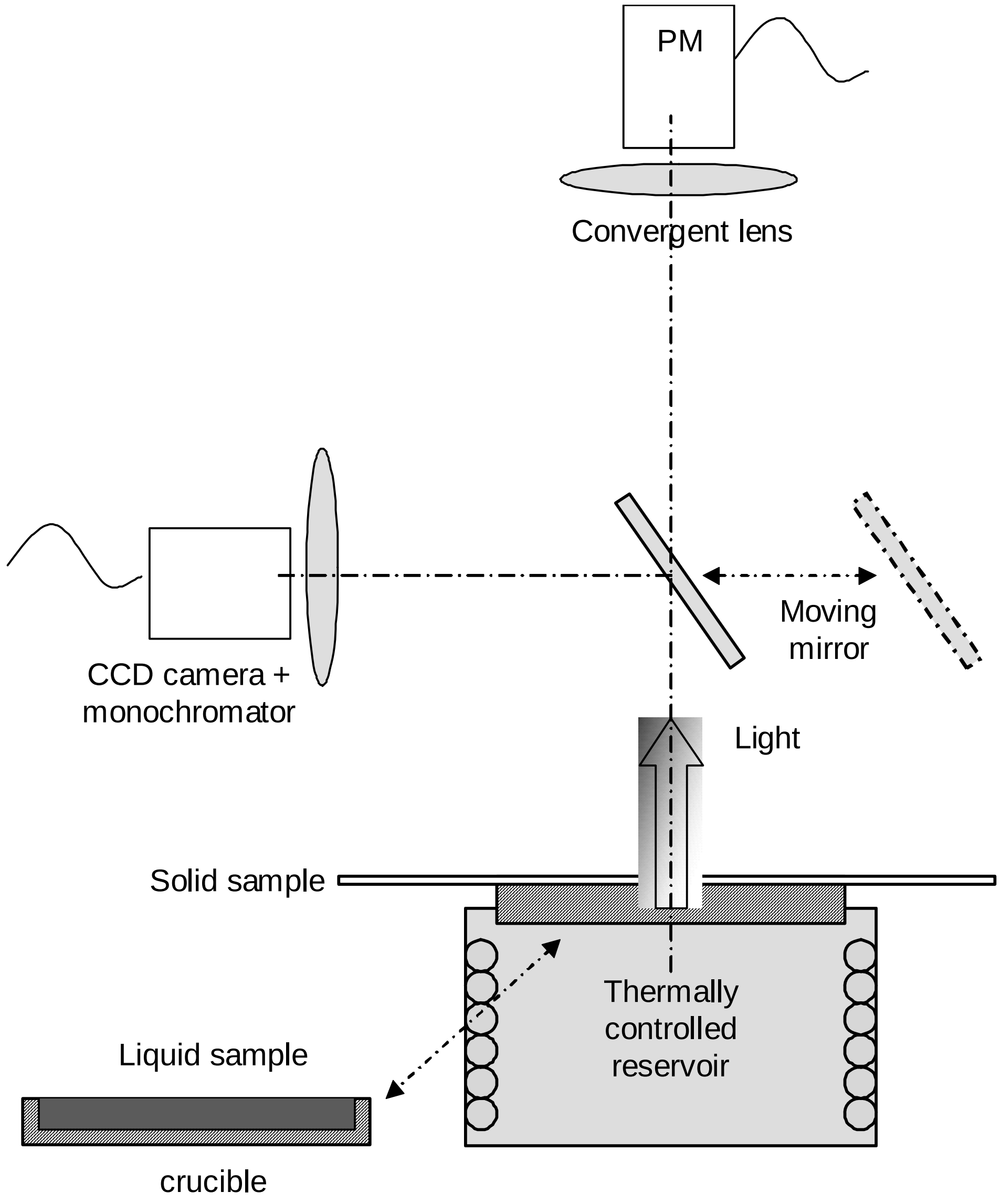

Figure 1

*Schematic of the experimental set-up for light detection*

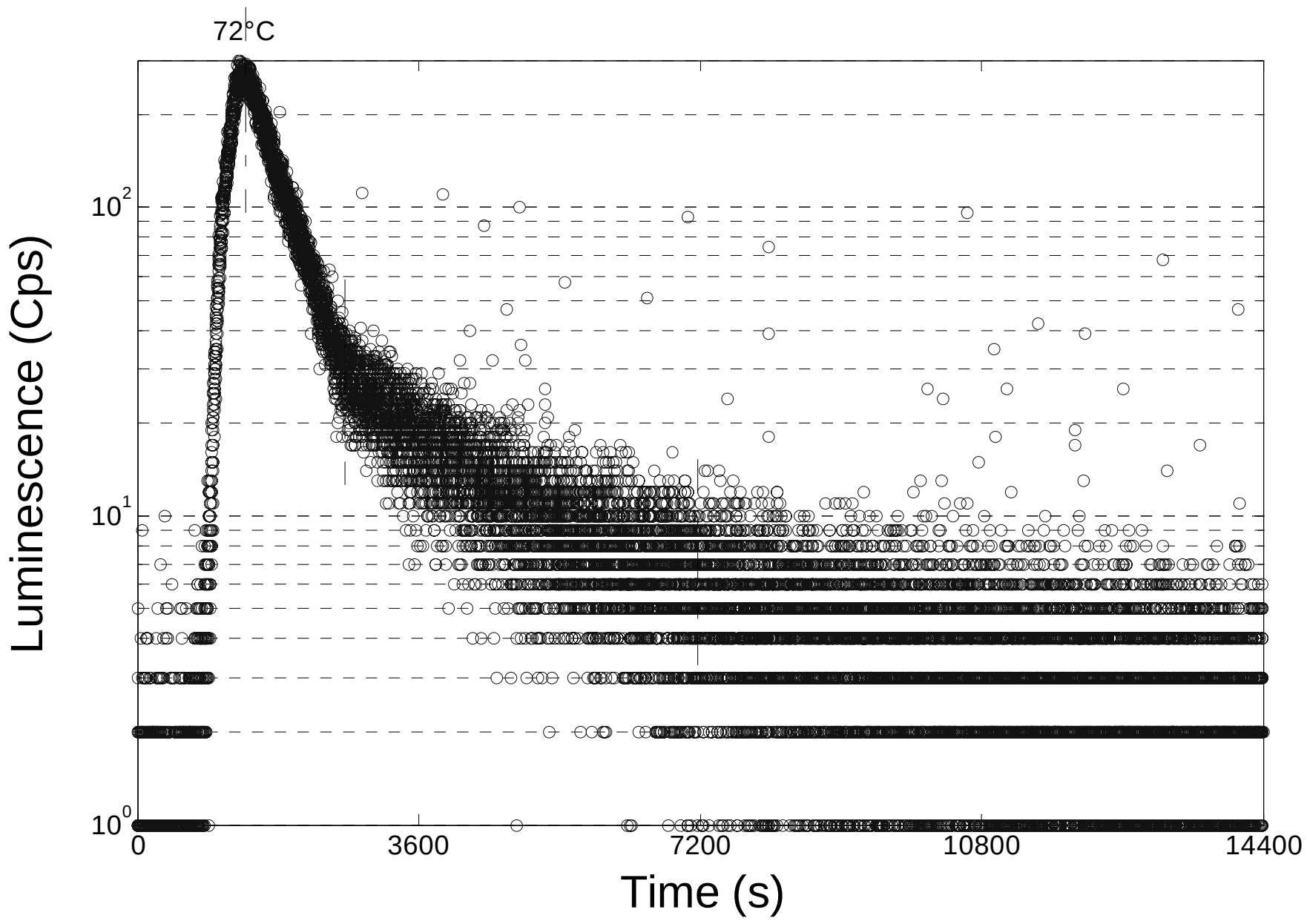

Figure 2

*Luminescence kinetics measured on a cured sample under a nitrogen atmosphere (temperature rise of 5°C/min up to 72 °C). Cps stands for photomultiplier counts per second.*

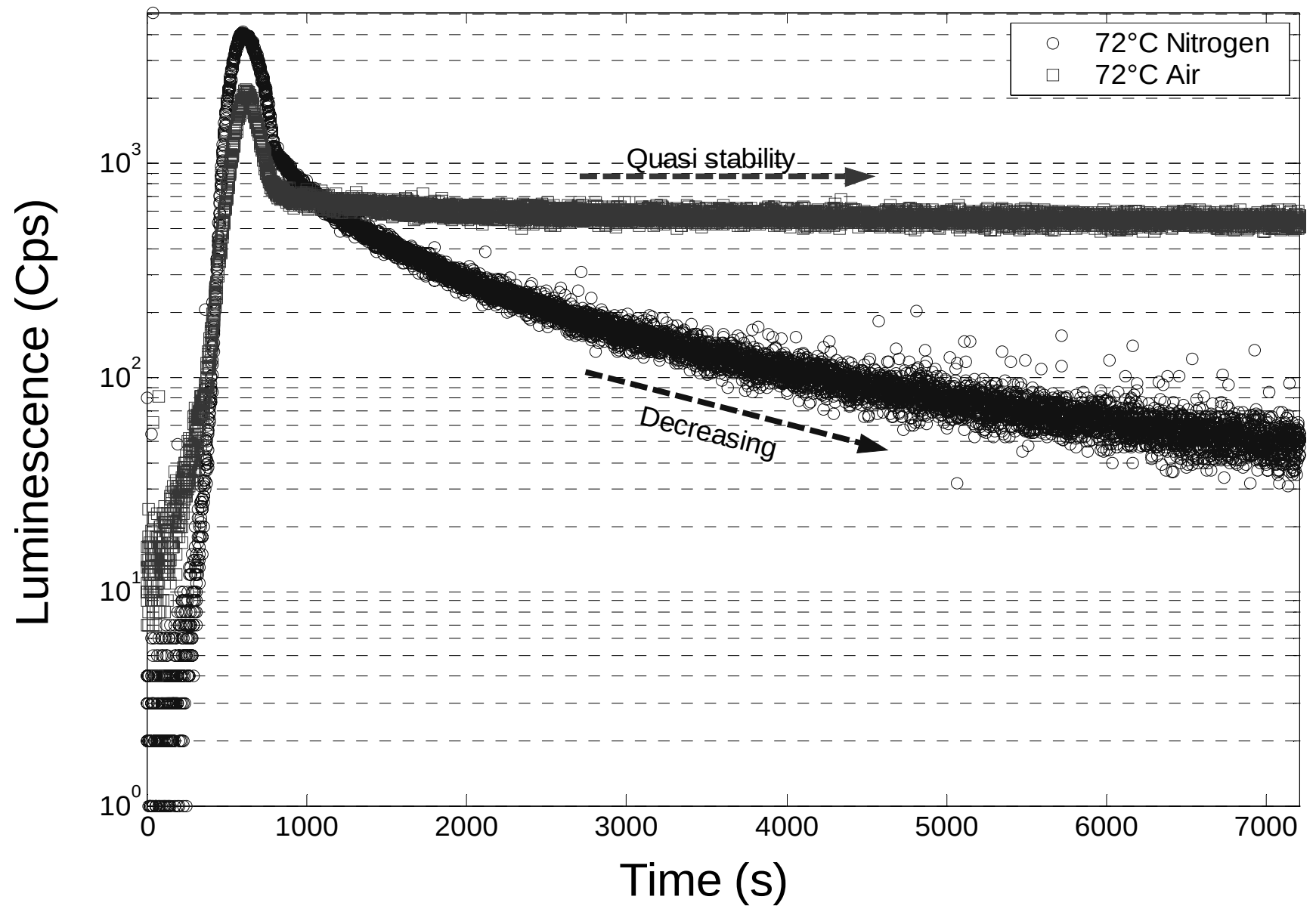

Figure 3

*Comparison between the luminescence kinetics measured on a cured sample under both nitrogen and air atmospheres (temperature rise of 5°C/min up to 72 °C)*

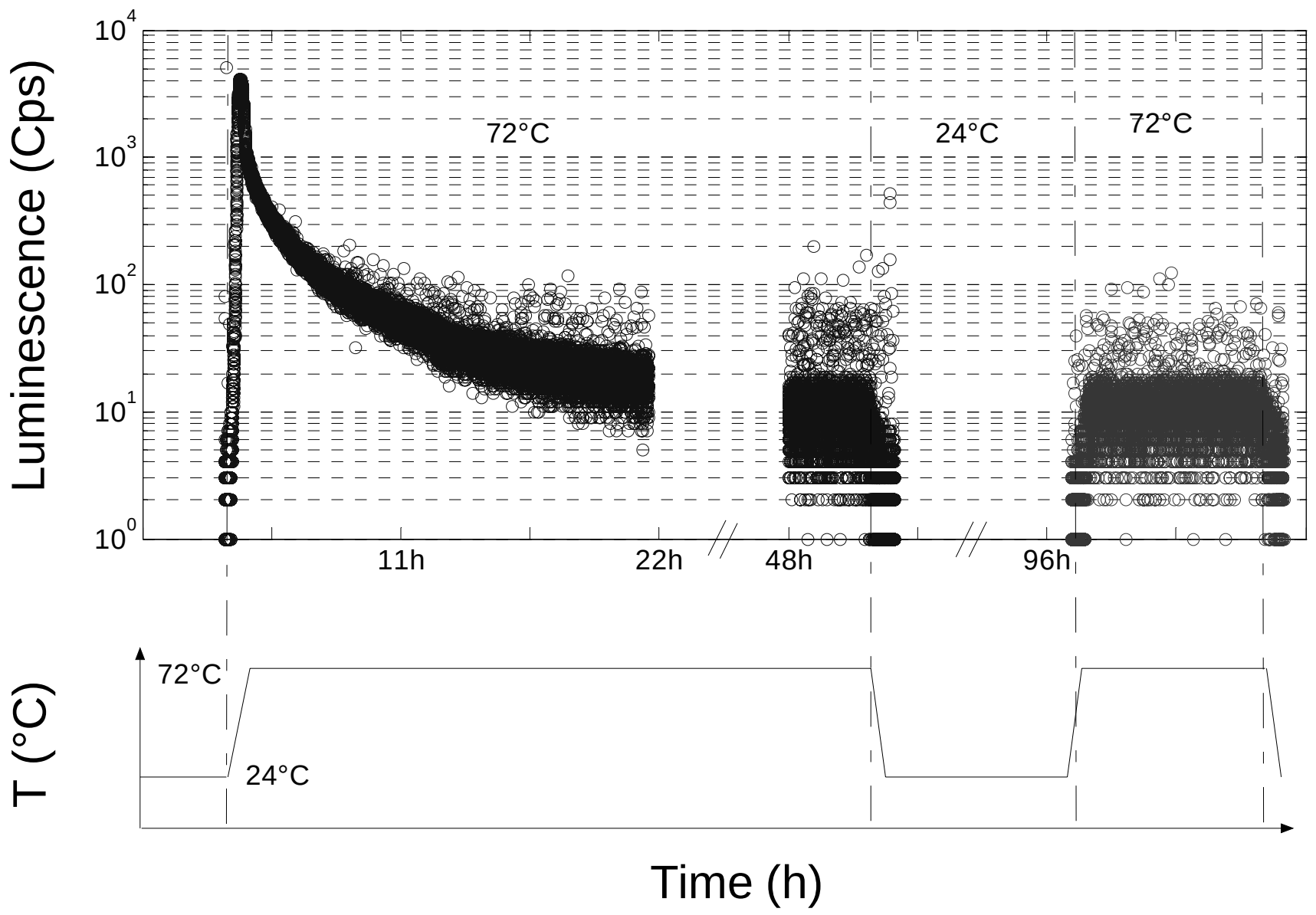

Figure 4

*Effect of temperature cycling on the luminescence emitted by a cured resin in nitrogen (rate of temperature change is 5°C/min).*

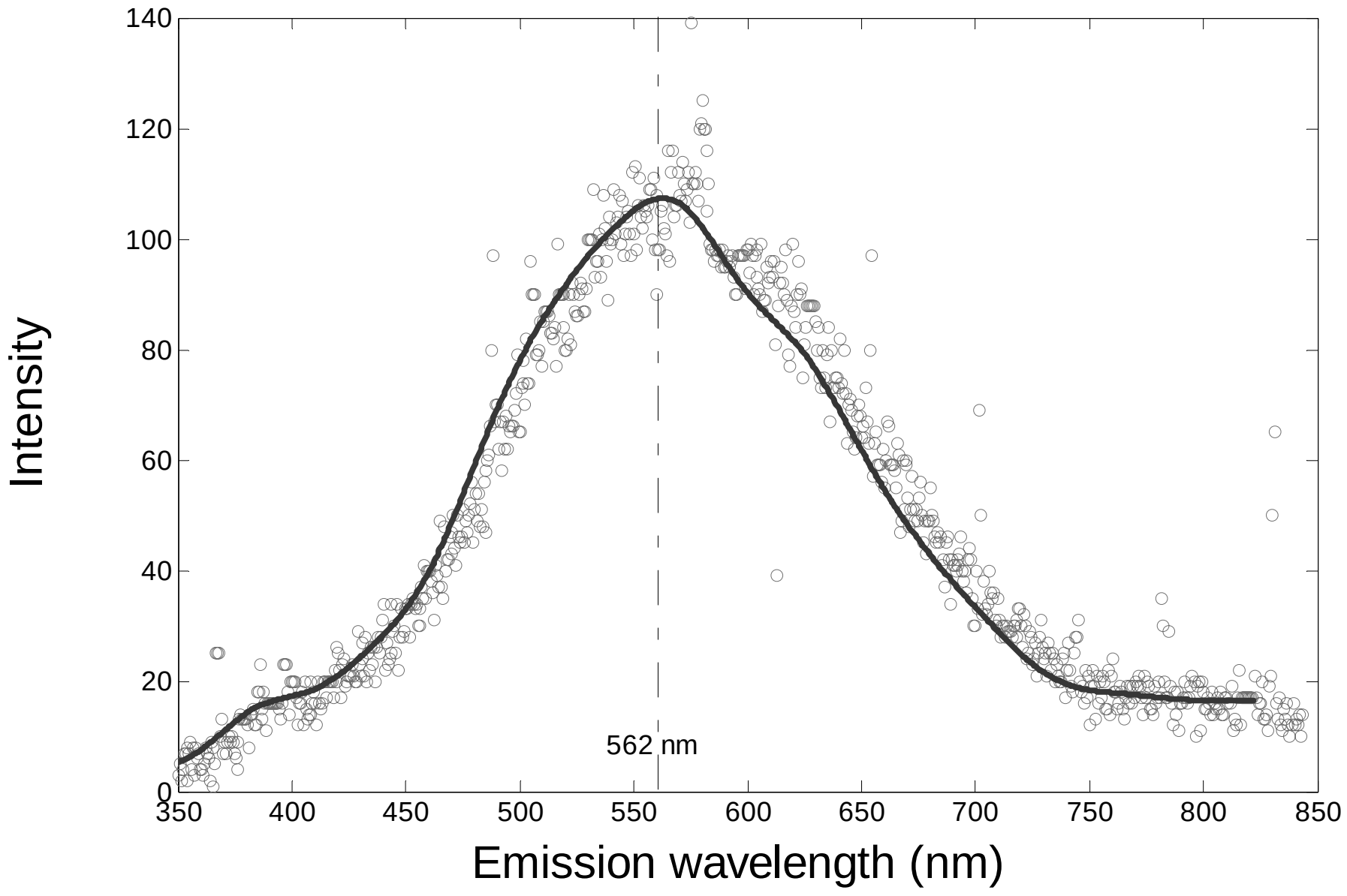

Figure 5

*Luminescence spectrum of cured resin at 100°C*

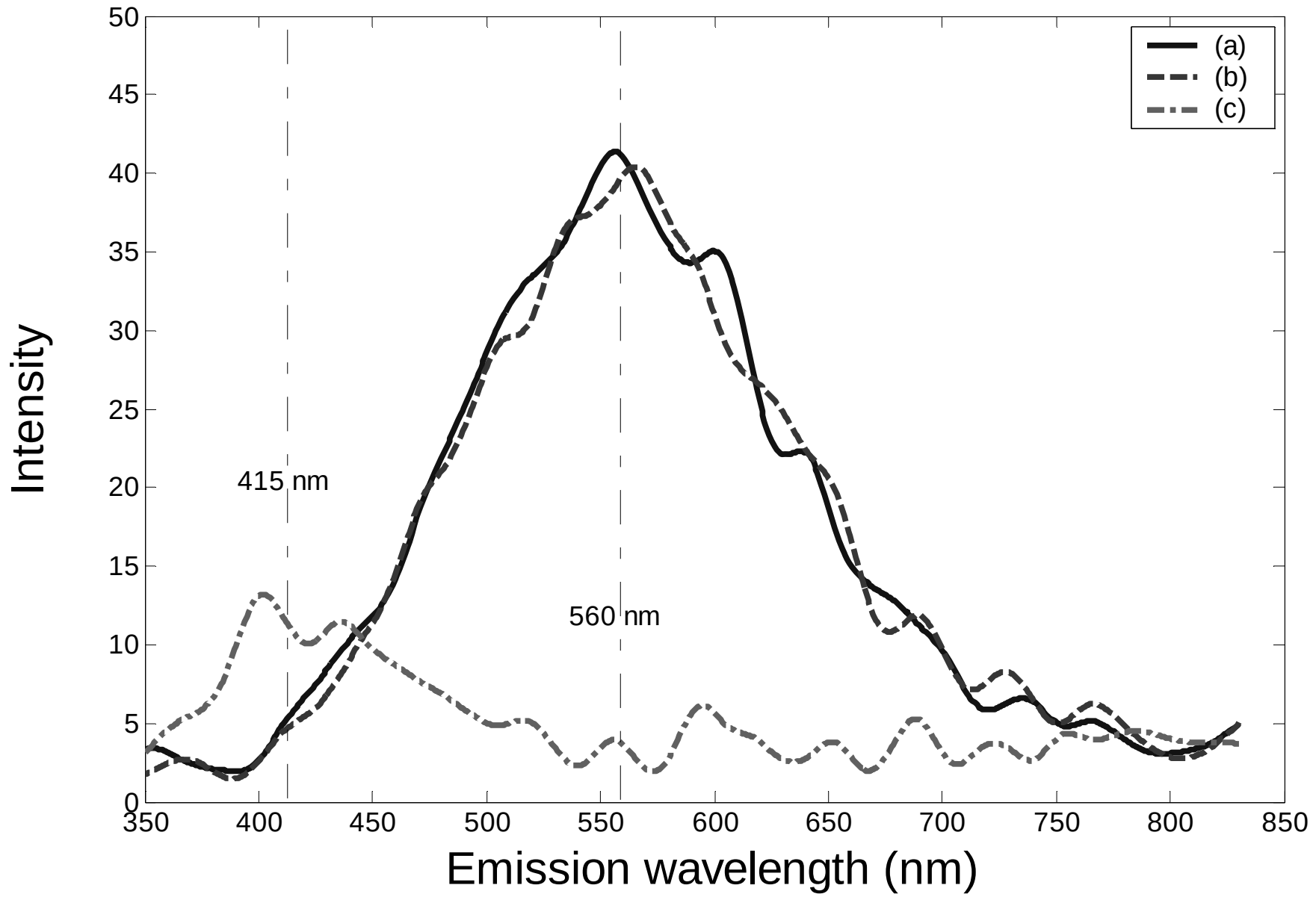

Figure 6

*Luminescence spectra after 30 min at 100°C: (a)-resin-hardener mixture during curing, (b)-hardener, (c)-base resin.*